# Improved energy confinement with nonlinear isotope effects in magnetically confined plasmas

J. Garcia[1], T. Görler[2], F. Jenko[3]

[1]*CEA, IRFM, F-13108 Saint-Paul-lez-Durance, France.*

[2] *Max Planck Institute for Plasma Physics, Boltzmannstr. 2, 85748 Garching, Germany*

[3]*Department of Physics and Astronomy, University of California, Los Angeles, California 90095, USA*

The efficient production of electricity from nuclear fusion in magnetically confined plasmas relies on a good confinement of the thermal energy. For more than thirty years, the observation that such confinement depends on the mass of the plasma isotope and its interaction with apparently unrelated plasma conditions has remained largely unexplained and it has become one of the main unsolved issues. By means of numerical studies based on the gyrokinetic theory, we quantitatively show how the plasma microturbulence depends on the isotope mass through nonlinear multiscale microturbulence effects involving the interplay between zonal flows, electromagnetic effects and the torque applied. This finding has crucial consequences for the design of future reactors since, in spite of the fact that they will be composed by multiple ion species, their extrapolation from present day experiments heavily relies on the knowledge obtained from a long experimental tradition based in single isotope plasmas.

The impact of the different isotopes of an element has been found to be important in a broad variety of physical systems. One example is the role of isotope effects on the critical temperature for the transition to the superconducting state of mercury [1]. Similar changes have been also found in magnetically confined plasmas, which are complex nonlinear physical systems whose main aim is to provide fusion power as a source of electricity. Such plasmas are generated and controlled in experimental devices producing specific magnetic configurations, e.g., tokamak or stellarator geometries [2]. The thermal energy confinement time has been observed to largely fluctuate with the exchange of the main isotope in broad experimental conditions [3]. Of particular interest is the isotope exchange between the fusion of deuterium (DD) and deuterium-tritium (DT) nuclei, as the first reaction is the one usually produced in present day tokamaks (where the main ion is D), whereas the second one produces much higher rates of fusion energy and it is envisaged to generate fusion power in future tokamaks as ITER [4]. In specific experiments performed in the decade of the 1990's both in the Tokamak Fusion Test Reactor (TFTR) and the Joint European Torus (JET) a broad an unexpected variety of results, including strong ion heat transport reductions, were obtained when changing the main ion in the plasma composition from D to a mixture of DT [5,6]. It was obtained that the plasma thermal energy confinement with different isotope configurations had a dependence on apparently unrelated physical variables as the magnetic geometry and the plasma beta, $\beta = 2\mu_0 <P>/B^2$, where P is the plasmas pressure and B the magnetic field or the input power. For more than thirty years these phenomena have not been explained, leading to the so-called isotope effect, which is a major drawback for the extrapolation of present-day results to future fusion tokamak devices as this extrapolation is usually performed based on the so-called Gyro-Bohm scaling [4], which assumes that the characteristic step size of collisional transport and turbulent structures both increase with the plasma gyroradius, $\rho = m_i v_i / |e| B$ with $m_i$ the ion mass, $v_i$ the perpendicular ion velocity, $e$ the electron mass and $B$ the magnetic field. Assuming that $v_i$ scales as the ion thermal velocity, $v_i = v_{th,i} = \sqrt{T_i/m_i}$ then one expects the ion heat flux to be $Q_i \sim \rho^2 v_{th,i}/L$ with $L$ a suitable macroscopic length, something that leads to a dependence $Q_i \sim m_i^{1/2}$ which means an increase by about 12% when the plasma mass increases from DD to DT, therefore in contradiction with many of the results obtained in isotope studies. However, the Gyro-Bohm scaling is a well stablished law which has been validated in dedicated scans of, for instance, $\rho^* = \rho/a$ with a the tokamak minor radius [7]. Indeed, a possible route to a complete understanding of the issue is assuming that the scaling is broken by some physical mechanism. It is well known that, at least for microturbulence driven by ion temperature gradients, referred to as the Ion Temperature Gradient (ITG) mode [8], which is responsible for the heat transport in the vast majority of the present day tokamaks, turbulent eddies can be quenched by the background ExB flow shear. The ExB growth rate is expected to be independent of the mass, $\gamma_{ExB} \sim E_r/L$ whereas the ITG growth rate scales as $\gamma_{ITG} \sim v_{th,i}/L$ and therefore the ratio $\gamma_{E \times B}/\gamma_{ITG}$, a measure of the impact of the external ExB flow shear on turbulence, scales as $\gamma_{E \times B}/\gamma_{ITG} \sim m_i^{1/2}$ indicating that the effectiveness of the ExB flow shear for quenching ITG transport increases with the mass at constant $\gamma_{ExB}$. However, the increased mass would also lead to increased Gyro-Bohm transport and therefore the final result cannot be derived from these simple considerations.

Additional arguments have been raised to explain the isotope effect, e.g. the change of the edge plasma conditions [9] and the different impact of zonal flows, i.e., large scale modes that are driven by turbulence and in return regulate turbulent transport on ITG turbulence [10,11]. However a full explanation of the isotope effect on plasma confinement, the linking with another plasma conditions and the consequences for future DT campaigns have been elusive.

In this paper, we quantitatively show for the first time, using the gyrokinetic theory [12], that the interplay between nonlinear microturbulence effects generates the isotope effect leading to an increase or decrease of ion heat fluxes from DD to DT plasmas in agreement with the trends found in previous experiments and providing clues about how to proceed in the future for maximizing thermal energy confinement in the presence of DT mixtures.

For that purpose and in order to properly quantify the implication that a change from DD to DT would have in future tokamaks devices with enough fusion power production, an ITER case has been selected [13]. It corresponds to a so called hybrid regime [14] with high β and input power, and a thermal energy enhancement factor relative to the IPB98(y,2) scaling [15], $H_{98}(y,2)$, higher than 1. The safety factor, q, defined as the ratio of the times a particular magnetic field line travels around the toroidal direction to the poloidal direction, is low with a weak magnetic shear. A summary of the main plasma parameters can be found in [13]. In this study a mixture of 50%D and 50%T has been assumed.

In order to analyze the impact of the isotope exchange on the ion heat flux, the gyrokinetic code GENE [16] has been used for computing linear and nonlinear microturbulence characteristics in the core plasma region. All simulations included kinetic electrons, collisions and electromagnetic effects. The geometry used was calculated by the code HELENA [17] which solves the Grad-Shafranov equation. The instability linear growth rates γ and the external ExB flow shear rate, $\gamma_{ExB,ext} = r/q \, d\Omega/dr$, with Ω the toroidal angular velocity, are in units of $C_s/R$, with $C_s = \sqrt{T_e/m_i}$ and $T_e$ the electron temperature, $m_i$ the deuterium mass and R the tokamak major radius. Both $\delta B_\perp$ and $\delta B_\parallel$ fluctuations were computed as they can both play a significant role in high β discharges. Recently it was found that a high fraction of energetic ions can change microturbulence through the pressure gradients in electromagnetic gyrokinetic simulations [18,19]. This effect can be important for the comparison of DD and DT plasmas and therefore the energetic ion distribution has been taken into account in the DT simulation by including two extra ion species owing to the Neutral Beam Injection (NBI) heating and the alpha particles produced in the DT reactions. The fast ion distribution function was approximated as Maxwellian, taking the average energy of the fast ion slowing-down distribution obtained in [13]. For the gyrokinetic calculations excluding the energetic ion population, which are performed with the aim of just analyzing the impact of exchanging D by T without any extra contribution, the equilibrium was recalculated with only the thermal pressure, and then used for the simulations. The dimensionless parameters of the discharge at $\rho = 0.33$, with $\rho = \sqrt{\varphi}$ and $\varphi$ the normalized toroidal flux, fed into the gyrokinetic calculations are summarized in Table I.

| | s | q | $T_e/T_i$ | $R/L_{Ti}$ | $R/L_{Te}$ | $R/L_{Ne}$ | $T_{f,beams}/T_e$ | $R/L_{Tf,beams}$ | $R/L_{Nf,beams}$ | $T_{f,\alpha}/T_e$ | $R/L_{Tf,\alpha}$ | $R/L_{Nf,\alpha}$ |
|---|---|---|---|---|---|---|---|---|---|---|---|---|
| ITER | 0.24 | 1.17 | 1.09 | 4.1 | 2.9 | 1.9 | 22.5 | 1.85 | 13.1 | 41.3 | 0.94 | 9.23 |

TABLE I. Discharge dimensionless parameters at $\rho = 0.33$ used as input in simulations

At first, a linear analysis has been carried out for different plasma configurations. In figure 1a and 1b, the linear growth rates and frequencies are shown. Over a wide normalized wavenumber range, $k_y = \hat{k}_y \rho_s$ where $\rho_s$ is the ion gyroradius with respect to the sound speed, ITG modes are unstable except for the full DT standard case including the fast ion population for which a high frequency mode appears at $k_y=0.05$. This mode is identified here as Alfven Beta Eigenmode (BAE)/Kinetic Ballooning Modes (KBM) [20], mainly driven by the energetic particles. Therefore the simulation lies just at the boundary between ITG and KBM regimes as obtained in hybrid plasmas from JET [19]. From the simulations without the energetic ion population, it is clear that the maximum growth rate, $\gamma_{max}$, is lower for DT, $\gamma_{max,DT} = 0.034 \, [C_S/R]$, than for DD, $\gamma_{max,DD} = 0.042 \, [C_S/R]$, however this trend closely follows Gyro-Bohm scaling as $\gamma_{max,DT} \sim \gamma_{max,DD}\sqrt{m_{DD}/m_{DT}}$ something expected from the fact that the ITG growth rate scales with the thermal ion velocity $v_{th,i} = \sqrt{T_i/m_i}$. Following the scaling of the gyroradius with mass there is also a shift towards lower values of the $k_{y,max}$ corresponding to $\gamma_{max}$ for

DT. In both cases, $\gamma_{max}$ is higher compared to the case including energetic particles, reflecting the fact that such particles can play a role in fusion reactor conditions. The influence of electromagnetic effects was studied performing the same type of simulation by reducing the electron beta $\beta_e = 2\mu_0 <P_e>/B^2$ from the nominal value 1.2% to $1.2\times10^{-2}$ % (i.e. performing an electrostatic simulation). The growth rates increase for both DD and DT however the relative difference between both remains similar to the electromagnetic case. Therefore, from the linear analysis, no significant deviation from the expected Gyro-Bohm scaling is obtained.

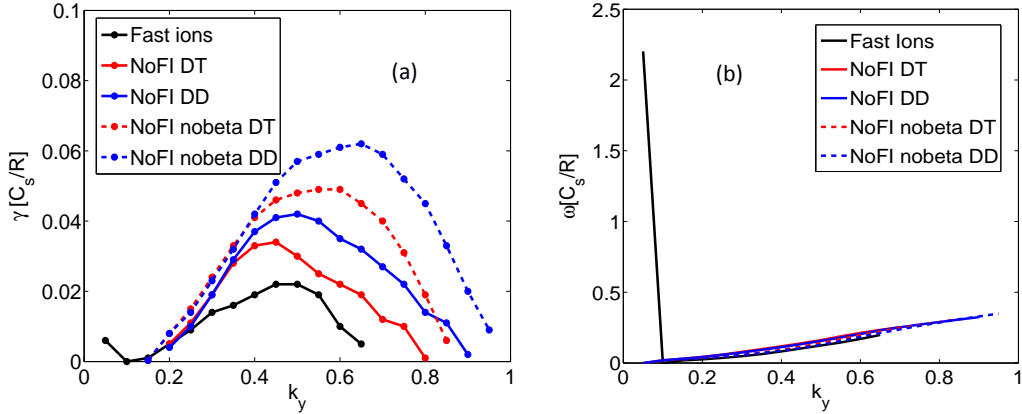

Figure 1 (a) Linear growth rates for the case including the fast ion content in DT, excluding the fast ion content in DT, excluding the fast ion content in DD, excluding the fast ion content and electromagnetic effects in DT and excluding the fast ion content and electromagnetic effects in DD (b) Frequency for the previous cases.

Non-linear simulations have been also performed for the previous linear runs including as well the effect of the external ExB flow shear, $\gamma_{ExB,ext}$, which is known to quench turbulence in the ITG domain [21] and the consequent parallel flow shear (pfs) which can destabilize it [22]. Due to the uncertainties on the external torque on ITER, which is the main source of $\gamma_{ExB,ext}$, there is no consensus of the value for this parameter, however it is expected to be considerably lower than in present day tokamaks due to the low torque applied. Therefore, a reduced value, in agreement with ITER integrated modelling simulations [23], has been assumed, $\gamma_{ExB,ext}$=0.01. A summary of the heat fluxes together with $\gamma_{max}$ and $k_{y,max}$ is shown in table II. Additionally, the influence of the zonal flow activity, which tend to squeeze and break the turbulent eddies, regulating the turbulence level and transferring energy to smaller spatial scales, is evaluated with the quotient of the average zonal ExB shear rate, $\gamma_{ExB,zonal}$, and the maximum linear growth rate. This represents the ratio between the strength of the instability (growth rate) and the intensity of one of the lead saturation mechanisms. The shear rate is defined as $\gamma_{ExB,zonal} = \frac{\partial}{\partial r}\langle v_{E\times B}\rangle$. The values of $\gamma_{ExB,zonal}/\gamma_{max}$ are also shown in table II for the different plasma conditions assumed.

| # | Case | $Q_i$(MW/m²) | $\gamma_{max}$[Cs/R] | $k_{y,max}$ | $\gamma_{ExB,zonal}$[Cs/R] | $\left|\gamma/k_y^2\right|_{max}$ | $\gamma_{ExB,zonal}/\gamma_{max}$ |
|---|---|---|---|---|---|---|---|
| 1 | DT + fast ions | 84 | | | 0.34 | | |
| 2 | DT | 154 | | | 0.37 | | |
| 3 | DD | 271 | | | 0.42 | | |
| 4 | DT no ExB | 308 | 0.034 | 0.45 | 0.43 | 0.228 | 12.65 |
| 5 | DD no ExB | 363 | 0.042 | 0.50 | 0.45 | 0.238 | 10.71 |
| 6 | DT electrostatic | 1270 | | | 0.72 | | |
| 7 | DD electrostatic | 1505 | | | 0.72 | | |
| 8 | DT no effect | 1491 | 0.050 | 0.60 | 0.69 | 0.269 | 14.08 |
| 9 | DD no effect | 1366 | 0.062 | 0.65 | 0.68 | 0.262 | 10.48 |

TABLE II. Results obtained for the different cases considered in this paper.

The total ion heat flux has its minimum for the case 1 as the increased energetic ion pressure (mainly due to the fusion reactions) suppresses turbulence. However, surprisingly, just by changing the isotope from DD to DT the heat flux is reduced by 43% from $Q_{i,DD} = 271\,kW/m^2$ to $Q_{i,DT} = 154\,kW/m^2$ as shown for cases 2 and 3. This trend is indeed not following the Gyro-Bohm scaling. On the other hand, when no electromagnetic or external ExB flow shear effects are included, cases 8 and 9, the heat flux obtained for DT is higher than DD and actually it closely follows the Gyro-Bohm scaling as $Q_{i,DT}/Q_{i,DD} = 1.09 \sim \sqrt{5/4}$, in agreement with previous studies in electrostatic conditions [24] and the ordering of heat transport as $Q_i \sim \rho^2 v_{th,i}/L$ . It is worth pointing out that the heat fluxes obtained in this study are close to the one obtained in the original ITER simulation in Ref. [13], $Q_i = 134\,kW/m^2$.

In order to identify the physical mechanisms which cause the plasma turbulence to deviate from such scaling (and even reducing the heat flux) the same simulations have been performed by removing the ExB external flow shear, performing electrostatic simulations and removing both effects at the same time. The results with $\gamma_{ExB,ext}$=0 (cases 4 and 5) but including electromagnetic effects show that the strong heat flux difference between DD and DT is now reduced to just 15% with $Q_{i,DD} = 363\,kW/m^2$ and $Q_{i,DT} = 308\,kW/m^2$. A plausible explanation for the higher impact of $\gamma_{ExB,ext}$ on DT comes from the linear analysis previously performed. The fact that $\gamma_{max}$ is lower for DT makes the ratio $\gamma_{ExB,ext}/\gamma_{max}$ higher. Additionally, the fact that $k_{y,max}$ is lower makes its efficiency higher. More difficult to interpret is the striking fact that a similar trend is found when performing electrostatic simulations and still including external ExB flow shear, cases 6 and 7. In this case, $Q_{i,DD} = 1505\,kW/m^2$ and $Q_{i,DT} = 1270\,kW/m^2$ and therefore the difference reduces to 15.6%, similar to the case without external ExB. It is worth to point out that this strong impact of the electromagnetic effects on the isotope effect is a pure non-linear effect as the linear analysis showed that $\gamma_{max}$ closely follows Gyro-Bohm scaling. Of particular interest is whether the so called quasi-linear models for the heat flux, which usually use the weight $q_l = \left|\gamma/k_y^2\right|_{max}$, where the maximum is taken over the k-spectrum, as a mixing length rule for the calculation of the heat transport, can reproduce the trend found in this paper in the case of Gyro-

Bohm breaking due to electromagnetic effects. As shown in table II, $q_l$ is able to capture the reversal of the heat flux from the cases 8 and 9 to the cases 4 and 5 when the electromagnetic effects are taken into account, however, the magnitude is much weaker than the full non-linear simulations. It is worth to clarify that the $k_y$ corresponding to $|\gamma/k_y^2|_{max}$ is $k_y=0.35$ for all the cases.

In order to verify whether the impact of zonal flows can explain the Gyro-Bohm breaking for the electromagnetic case, an analysis is performed here by calculating the ratio $\gamma_{ExB,zonal}/\gamma_{max}$. Comparing the DT and DD cases at high (4 and 5) and low (8 and 9) β, shows that the impact of zonal flow is indeed stronger for DT than DD. However there is no direct translation of this trend to the heat fluxes, as for the cases 8 and 9 the heat flux is higher for DT in spite of the fact that $\gamma_{ExB,zonal}/\gamma_{max}$ is also higher. Therefore this explanation is incomplete and cannot account for the phenomenology found in this study.

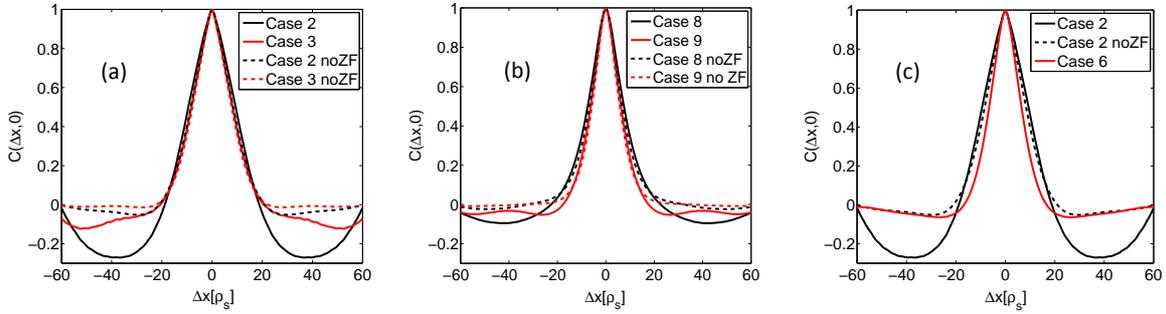

Figure 2 (a) Electrostatic potential equal-time two point correlation function as a function of the radial distance $\Delta x$ in $\rho_s$ units for DT and DD plasma including $\gamma_{ExB,ext}$ and electromagnetic effects (cases 2 and 3). The zonal flow component is removed in the post-processing analysis (b) Same analysis excluding $\gamma_{ExB,ext}$ and electromagnetic effects (cases 8 and 9). (c) Impact on the electrostatic potential equal-time two point correlation function of the electromagnetic effects in a DT plasma

One important point in the Gyro-Bohm scaling and in general in confined plasmas turbulence is the consideration that the heat transport turbulent structures characteristic size is just a few gyroradii. This implies that the heat transport is basically a local feature. In order to verify this hypothesis in the simulations performed here, the electrostatic potential, $\emptyset$, equal-time two point correlation function,

$$C(x, \theta = 0, \Delta x) = \frac{\langle \emptyset(x, \theta = 0)\emptyset(x + \Delta x, \theta = 0)\rangle}{(\langle \emptyset(x, \theta = 0)^2\rangle\langle \emptyset(x + \Delta x, \theta = 0)^2\rangle)^{1/2}} \quad (1)$$

where $\theta$ is the poloidal angle and $\langle \ \rangle$ the time calculated over the saturated part of the simulation, has been analyzed with the aim of having a deeper insight on the physical mechanisms playing a role in the isotope effect. The correlation length is defined as $C(x, \theta = 0, L_C) = 1/2$. Surprisingly, $L_C$ increases following Gyro-Bohm scaling from case 3 with DD, $L_C = 10.3\rho_s$ up to $L_C = 11.5\rho_s$ for the case 2 in spite of the fact that the heat flux reduces with DT. This trend has been recently confirmed in the tokamak ASDEX Upgrade when performing ρ* scaling experiments with hydrogen and deuterium as the main plasma gases [25], indicating a possible common origin for the isotope effect in single and multi-ion plasmas. On the other hand, for the cases 8 and 9, the heat flux and the correlation length now follow a similar trend from DD, $L_C = 8.03\rho_s$, to DT, $L_C = 9.44\rho_s$. This apparent contradiction can be understood by analyzing the full correlation function, i.e. analyzing long range correlations, shown in figure 2a for both cases, including and excluding the zonal flow component in the post processing analysis. There is a strong anticorrelation region for high $\Delta x > 20\rho_s$ for case 2 which is much weaker for case 3. The origin of this region is the high contribution of the zonal flow component for the DT mixture compared to DD. The same comparison is made for cases 8 and 9 when all the effects are suppressed, shown in figure 2b. The anticorrelation region is very much reduced for both cases although it is still evident that the zonal flow component is stronger for DT. From this analysis it is apparent that the stronger zonal flow for DT cannot compensate for the increased correlation length obtained from DD to DT and the general flux follows Gyro-Bohm scaling. Additionally, the role of electromagnetic effects becomes crucial since, as shown in figure 2c, the strong zonal flow activity, detected from the anticorrelation function for case 2, is mostly suppressed when the electromagnetic effects are not taken into account. Therefore, the Gyro-Bohm scaling, present at short scale length in any plasma condition, can be counteracted in DT by the long correlation zonal flow activity when the plasma β is high enough.

The previous findings suggest a solid route for improved confinement with respect to DD plasmas and can provide an optimum operational regime. High power (favoring the increase of beta, external torque and fast ions fraction) is recommended for strong isotope effect whereas for low power (and low torque) the confinement could be worse for DT than DD. It is worth to point out that the previous findings are in agreement with experimental evidence found in TFTR and JET. It was found that the confinement in DT increased with respect to DD with increasing power. In particular, it was found in TFTR [5] that, at constant high beta, up to three times less ion heat flux was obtained in the core with DT. The reduction of heat flux obtained in this paper is of that order as shown for the cases 1 and 3.

These results have critical impact on ITER and future tokamak reactors as the transition from the L-mode (with low confinement) to the H-mode (with higher confinement) [2] will be performed when the pressure, and consequently beta, will be still low [13]. Therefore, the plasma could not benefit of an improved confinement which would be favorable for the fusion power generation which in turn would positively strengthen the effect in a virtuous loop.

Finally, these findings can lead to a better understating of the link between isotope effects in apparently unrelated physical systems for which the magnetic field plays an important role, like paramagnetic-diamagnetic phase-transitions in superconducting materials, which also show an isotope effect [1]. Surprisingly, this kind of phase transition has been also identified in magnetically confined plasmas during the process of fully stabilization of turbulence [26,27] and, therefore, the possibility of a universal explanation for both will be explored in the future.

In conclusion, the here identified isotope effect, by which the thermal energy confinement changes in magnetic confined plasmas with the hydrogen isotope, can be understood as a complex nonlinear multiscale interaction involving external ExB flow shear, zonal flow activity and electromagnetic effects. The inherent Gyro-Bohm scaling for plasma microturbulence, which exists in any plasma condition and increases the radial correlation length at short scales, is highly counteracted by the concomitant appearance of these effects, leading to a broad variety of different types of confinement depending on the plasma conditions. Here gyrokinetic theory can provide an optimum plasma configuration for DT operational regimes, favoring high beta and input power with enough torque and fast ion content, characteristic of the so called advanced tokamak regimes [26,28]. The preparation and design of future DT campaigns like the one envisaged for JET [29] and tokamak reactors can indeed benefit from these findings. However, the final impact of these effects in DT plasmas would require detailed integrated modelling, including self-consistent sources for heating, fueling and torque, a work which will be carried out in the future.

**Acknowledgements**. The simulations presented in this work were carried out using the HELIOS supercomputer system at Computational Simulation Centre of International Fusion Energy Research Centre (IFERC-CSC), Aomori, Japan.